\documentclass[fleqn,usenatbib]{mnras}

\usepackage{newtxtext,newtxmath}
\usepackage{comment}

\usepackage[T1]{fontenc}

\DeclareRobustCommand{\VAN}[3]{#2}
\let\VANthebibliography\thebibliography
\def\thebibliography{\DeclareRobustCommand{\VAN}[3]{##3}\VANthebibliography}

\usepackage{graphicx}	
\usepackage{amsmath}	
\usepackage{float}
\usepackage{lipsum}
\usepackage{multirow}
\usepackage{xcolor}
\usepackage{orcidlink}

\title[Galaxy clustering at cosmic dawn]{Galaxy clustering at cosmic dawn from JWST/NIRCam observations to redshift z$\sim$11}

\author[Nicolò Dalmasso et al.]{
Nicolò Dalmasso$^{1,2}$\thanks{e-mail: ndalmasso@student.unimelb.edu.au}\orcidlink{0000-0002-1850-4050},
Nicha Leethochawalit$^{3}$\orcidlink{0000-0003-4570-3159},
Michele Trenti$^{1,2}$,
Kristan Boyett$^{1,2}$
\\
$^{1}$School of Physics, University of Melbourne, Parkville, Vic 3010, Australia\\
$^{2}$ARC Centre of Excellence for All Sky Astrophysics in 3 Dimensions (ASTRO 3D), Australia\\
$^{3}$National Astronomical Research Institute of Thailand (NARIT), Mae Rim, Chiang Mai, 50180, Thailand\\
}

\date{Accepted XXX. Received YYY; in original form ZZZ}

\pubyear{2024}

\begin{document}
\label{firstpage}
\pagerange{\pageref{firstpage}--\pageref{lastpage}}
\maketitle

\begin{abstract}
We report measurements of the galaxy two-point correlation function at cosmic dawn, using photometrically-selected sources from the JWST Advanced Deep Extragalactic Survey (JADES). The JWST/NIRCam dataset comprises approximately $N_g \simeq 7000$ photometrically-selected Lyman Break Galaxies (LBGs), spanning in the redshift range $5\leq z<11$. The primary objective of this study is to extend clustering measurements beyond redshift $z>10$, finding a galaxy bias $b=9.6\pm1.7$ for the sample at $\overline{z} = 10.6$. The result suggests that the observed sources are hosted by dark matter halos of approximately $M_{h}\sim 10^{10.6}~\mathrm{M_{\odot}}$, in broad agreement with theoretical and numerical modelling of early galaxy formation during the epoch of reionization. Furthermore, the JWST JADES dataset enables an unprecedented investigation of clustering of dwarf galaxies two orders of magnitude fainter than the characteristic $L_*$ luminosity (i.e. with $M_{F200W}\simeq-15.8$) during the late stages of the epoch of reionization at $z\sim 6$. By analyzing clustering as a function of luminosity, we find that $b(M_{F200W})$ aligns with previous results for brighter galaxies and then decreases with $M_{F200W}$, as theoretically expected for fainter candidates. These initial results demonstrate the potential for further quantitative characterisation of the interplay between assembly of dark matter and light during cosmic dawn that the growing samples of JWST observations are enabling. 
\end{abstract}




\section{Introduction}\label{sec: introduction}

\indent The Universe we observe today formed from primordial fluctuations emerged during an inflationary epoch, subsequently amplified by gravitational instability shaping matter distribution. Gas dissipated over time, gravitating towards dark matter halos' central regions, intricately linking the growth and spatial arrangement of galaxies to their dark matter halos (\citealt{Peebles_1980, Persic_1992, Bullock_2001}).\\
\indent The conceptualization of the galaxy–halo connection coincided with the realization that the spatial distribution of galaxies can yield insights into their formative properties. Early studies such as \citet{Peebles_1980} focused on the two-point statistics of galaxies, as well as investigations on signal strength sensitivity to measure galaxy mass and luminosity (e.g., \citealt{Bahcall_1983,Davis_1983}).\\
\indent Recognizing that measuring these clustering properties could furnish information about the masses of the dark matter halos housing these galaxies was a pivotal insight due to the pronounced dependence of halo clustering on halo mass (e.g., \citealt{Bardeen_1986,Mo_1996,Klypin_1996,Tormen_1999,Jenkins_2001,Tinker_2010}).\\
\indent A key tool to quantify clustering is the two-point correlation function ($\xi(r)$; \citealt{Peebles_1980}), which quantifies the excess probability of finding a galaxy at a given separation distance "r" compared to a random distribution. Given that precise redshift are not always available (e.g. in the case of photometrically selected samples), this function is often replaced by the Angular Two-Point Correlation Function (ACF), which expresses the excess as a function of the angular separation denoted by ($\theta$) instead. The ACF can then be mapped to an two-point correlation function in physical space based on the (estimated) redshift distribution of the sample, leading to a characteristic correlation length ($r_0$; \citealt{Davis_1983}). This parameter $r_0$ represents the typical separation distance at which galaxies exhibit significant correlation or clustering. The value of $r_0$ may vary depending on the specific sample of galaxies, the redshift range under investigation, and the cosmological model applied.\\
\indent The correlation length is closely tied to determining another crucial parameter: galaxy bias ($b$; \citealt{Peebles_1980}), representing the systematic deviation or clustering of galaxies compared to the underlying matter distribution in the Universe. A rising trend in galaxy bias suggests stronger clustering and in-turn this implies that the galaxy population is hosted in more massive dark-matter halos.\\
\indent Clustering, when combined with information on the galaxy luminosity function, can also provide insight on the typical occupation fraction of dark-matter halos that host luminous galaxies.This fraction is known as the duty cycle ($\epsilon_{DC}$). The duty cycle is typically derived by combining galaxy clustering measurements with a method called abundance matching (e.g., \citealt{Vale_2004}) by comparing the count of galaxies with luminosities surpassing a specified threshold, denoted as $L_g$, to the count of dark matter halos with masses exceeding a defined value $M_h$.\\
\indent Past studies established early in the history of high redshift galaxy observations that with higher UV luminosities tend to be hosted by more massive dark matter halos (e.g., \citealt{Giavalisco_2001, Foucaud_2003, Ouchi_2004, Ouchi_2005, Adelberger_2005,Lee_2006}). The results have continued to be confirmed as a well-established trend over the last decade (e.g., \citealt{Savoy_2011, Bian_2013, Barone_2014, Harikane_2016, Hatfield_2018, Qiu_2018, Harikane_2022, Dalmasso_2023}).\\
\indent Lyman Break Galaxies (LBGs) after the end of reionization at redshift $4.0<z<5.0$ show clear evidence for a luminosity-dependent bias, observed to the UV luminosity detection limit of past surveys $M_{UV}\lesssim-16.7$ (e.g., \citealt{Lee_2006, Harikane_2016, Harikane_2022}). This pattern persists when extending the redshift range to $5.0<z<6.0$, with absolute magnitudes ranging from $-22\lesssim M_{UV}\lesssim-18$. The findings range from $b=5.5^{+0.2}_{-0.4}$ for fainter objects to $b=7.7\pm0.1$ for brighter ones, further corroborating the trend of galaxy bias with luminosity (e.g., \citealt{Barone_2014, Harikane_2016, Qiu_2018, Hatfield_2018, Harikane_2022, Dalmasso_2023}). These findings align with expectations from basic theoretical modelling of the evolution of galaxy bias with luminosity based on assuming some efficiency in converting DM to stars, which may be calibrated at a reference redshift (e.g., \citealt{Trenti_2010,Trenti_2015}).\\
\indent Progress in clustering and bias measurement at high redshift is primarily limited by the challenges in building high quality samples of galaxies with robust redshifts. Furthermore, care must be applied when observations do not have uniform depth, as a spatially varying completeness can mimic a physical clustering signal. In \citet{Dalmasso_2023}, we introduced a novel approach to galaxy clustering measurements to address the issue of non-uniform catalogs in depth and completeness. These two traits stem from the fact that during the observation of a target sky area with the narrow fields of view of infrared cameras on Hubble (and JWST), mosaicing is needed. As a result certain regions are observed multiple times from overlapping pointings and dither patterns. This results in some areas having greater depth than others in the final survey mosaic (due to overlap of individual frames). In \citet{Dalmasso_2023} we demonstrated that these systematic sources of error can be effectively mitigated by generating a reference catalog of random points derived from artificial source recovery simulations. This allowed us to account effectively for map heterogeneities arising from multiple overlapping observations.\\
\indent With this method \citet{Dalmasso_2023} measured a galaxy bias of $b=9.3\pm4.9$ at redshift $z=7.7$ using the Hubble Space Telescope (HST) Legacy Field data (\citealt{Bouwens_2021}), which is the highest redshift clustering measurement reported so far. \\
\indent Thanks to the increased depth, spectral coverage and field of view of JWST/NIRCam compared to Hubble/WFC3, this study aims to extend the clustering redshift frontier further. In addition, we also aim to take advantage of the increased sensitivity to study the luminosity dependence of clustering reaching down to $M_{F200W}\lesssim -14.5$, approaching faint dwarf galaxies. To conduct this analysis, we utilize photometrically-selected sources from the JWST Advanced Deep Extragalactic Survey (JADES) detected by the NIRCam instrument.\\
\indent The structure of this paper is outlined as follows. In Sec.\ref{sec: data selection}, we elucidate our process for selecting data pertaining to the LBG candidates. In Sec.\ref{sec: clustering estimation}, we delve into the theory and variables estimation to study the ACF. Our discoveries and numerical results are expounded upon in Sec.\ref{sec: clustering results}. To conclude in Sec.\ref{sec: summary}, we offer a summary of our work and present our conclusions. In this work we assume, when relevant, the cosmological parameters determined by the Planck Collaboration \citep{ade2014planck}: $(\Omega_{M},\Omega_{\lambda}, h, \sigma_{8})=(0.315, 0.685, 0.673, 0.828)$. Magnitudes are in the AB system \citep{1983ApJ...266..713O}.

\section{Data selection}\label{sec: data selection}

This work is based on the NIRCam imaging public data release over the GOODS South field by the JWST Advanced Deep Extragalactic Survey (JADES)\footnote{\url{https://archive.stsci.edu/hlsp/jades}\label{note: jades}} collaboration (\citealt{Bunker_2023,Eisenstein_2023,Hainline_2023,Rieke_2023}).\\
\indent Specifically, in this study we utilized images obtained from JADES Version 1.0, acquired through the NIRCam instrument exhibiting a pixel scale of 30 milliarcseconds. As for the galaxies survey used in this work we considered the candidates detected in the F200W band presented in the "NIRCam Photometry"\footref{note: jades} catalog as well as the scientific images, segmentation maps, root mean square (rms) all presented in \citet{Rieke_2023}.\\
\indent For reference, we summarize briefly the detection and photometry methods used to create the catalog, referring the reader to \citet{Rieke_2023} for more details. The detection image was created by stacking signal and noise images of specific filters from the NIRCam LW. Python-based tools like \texttt{Astropy} \citep{Astropy_Collaboration_2022}, \texttt{Photutils} \citep{Bradley_2023}, \texttt{scikit-image} \citep{van_der_Walt_2014}, and \texttt{cupy} \citep{cupy_2017} were utilized, drawing inspiration from \texttt{NoiseChisel} \citep{Akhlaghi_2015}. A detection catalog was generated using the \texttt{Photutils detect-sources} routine, followed by operations to reduce noise and separate blended objects. Deblending was performed iteratively, and satellites were identified within extended sources. A final segmentation map was defined, incorporating masks for bright stars and persistence. The segmentation map, generated through the described detection method, served as the basis for constructing the JADES photometric catalog. Photometric redshifts were measured using the template-fitting code \texttt{EAZY} \citep{Brammer_2008} with the templates from \citet{Hainline_2023} and PSF-matched Kron fluxes.\\
\indent To create our parent catalog of galaxies we took the photometric catalog from \citet{Rieke_2023} and selected candidates within the redshift interval of $5.0\leq z<11.0$. This procedure resulted in the delineation of six discrete redshift bins, each characterized by an identical width of $\Delta z = 1.0$, along with an associated assignment of an average redshift value denoted as $\overline{z}$ for each bin namely $\overline{z}=$ 5.5, 6.5, 7.4, 8.5, 9.3, and 10.6. Second, within each of these redshift bins, we select candidates that exhibited a signal-to-noise ratio exceeding $SNR\geq 5.0$ in the UV band. Third, to enhance the precision of our analysis and identify sources in adjacent spatial regions, we ensure that the galaxies are confined to the observable region in both rms and segmentation maps.\\
\indent Lastly, capitalizing on the exceptional depth and sensitivity of JWST, we aimed to explore the cosmos at ever-diminishing magnitudes. Given that uncertainty in the estimation of completeness of a survey may affect the clustering estimation, especially if there are spatial variations across the field, we impose a magnitude threshold of $m_{F200W} < 31.1$ ($M_{F200W} < -15.5$ at $z\simeq5.5$). The corresponding completeness ranges from 8\% to 35\% for a magnitude range of $30.1 < m_{F200W} \leq 31.1$. Therefore, we opted to include only the sub-regions with a completeness above 30\%, ensuring a robust analysis in the faintest regime. For brighter objects, the completeness ranges from 60\% to 73\% for $28.6 < m_{F200W} \leq 30.1$, and from 73\% to 92\% for $26.6 < m_{F200W} \leq 28.6$.\\
\indent The outcome of the data selection process is displayed in the initial four columns of Tab.\ref{tab: clustering results}. These columns present the redshift bin (\#1), the mean redshift designated as $\overline{z}$ (\#2), the total count of galaxies within each redshift bin from the source catalog (\#3), and the number of galaxies we selected in each bin for subsequent analysis based on the selection criteria explained above (\#4).

\section{Clustering estimation}\label{sec: clustering estimation}

\subsection{ACF fitting model}\label{subsec: fitting model}

We start by constructing an estimator $w_{obs}$ for the observed ACF. This estimator measures the excess probability of encountering pairs of objects, specifically galaxies within our investigation, at a specific angular separation denoted as ($\theta$) compared to a uniform random distribution. We follow \citet{1993ApJ...412...64L}, which define $w_{obs}$ as follow:
\begin{equation}\label{eq: estimator}
  w_{\text{obs}}(\theta)= \frac{DD(\theta) - 2DR(\theta) + RR(\theta)}{RR(\theta)}
\end{equation}
In this equation, $DD(\theta)$ identifies the count of pairs of observed galaxies falling within an angular separation range of $(\theta \pm \delta\theta)$. $RR(\theta)$ corresponds to a similar count but is derived from a random catalog that shares identical geometric and selection characteristics as the observed catalog. $DR(\theta)$ denotes the count of pairs that include one observed galaxy and one object from the random catalog.\\
\indent The finite size of this survey results in an underestimation of the measured parameters, a bias that can be corrected by introducing a constant known as the integral constraint (IC; \citealt{10.1093/mnras/253.2.307}):
\begin{equation}
  w_\text{true}(\theta)=w_\text{obs}(\theta) + IC
\end{equation}
This correction constant is estimated as:
\begin{equation}\label{eq: IC}
\begin{split}
  IC &=\frac{1}{\Omega^{2}}\int_{1}\int_{2}w_\text{true}(\theta)d\Omega_{1}d\Omega_{2}\\ &=\frac{\sum_{i} RR(\theta_{i}) w_\text{true}(\theta)}{\sum_{i} RR(\theta_{i})} =\frac{\sum_{i} RR(\theta_{i}) A_{w}\theta^{-\beta}_{i}}{\sum_{i} RR(\theta_{i})}
\end{split}
\end{equation}
In the above equations, $w_\text{true}(\theta)$ represents the intrinsic ACF, $w_\text{obs}(\theta)$ is the measured ACF within the survey area and $RR(\theta_i)$ is the random-random count of galaxies in a specific angular bin.\\
\indent Owing to variations in the sizes of candidate subsamples within each redshift bin under examination in this analysis (refer to column 4 in Tab.\ref{tab: clustering results}), a decision was made to partition the investigated angular range, denoted by $\theta\in(0,250)"$, into varying numbers of angular bins tailored for each redshift range considered. The guiding criterion adopted involves subdividing the angular range into a finite set of bins, ensuring an approximately uniform count of candidate pairs ($DD$) per bin. This binning strategy was adopted to ensure that Poisson noise and sample sizes remain consistent across different angular separations. The results are then plotted on a logarithmic scale, but this has no impact in the determination of the bin sizes and ranges.\\
\indent Following best practices from prior work (e.g., \citealt{Lee_2006, Overzier_2006, Barone_2014, Dalmasso_2023}), we assume a fixed value of $\beta = 0.6$ and consider a maximum separation of $\theta=250"$. When the correlation slope $\beta$ is held constant, the quantity $IC/A_{w}$ is solely contingent on the size and configuration of the survey area, and it can be calculated using Eq.\ref{eq: IC}. The single parameter available for optimization is $A_w$, which is fine-tuned by minimizing the $\chi^2$ value. The fitting model is outlined as follows:
\begin{equation}\label{eq: estimator model}
  w_\text{mod}(\theta)=A_{w}(\theta^{-\beta} - IC/A_{w})
\end{equation}
\indent To assign uncertainties to the ACF measurements, we construct the normalized covariance matrix using the standard estimator:
\begin{equation}
C_{ij} = \frac{1}{N - 1} \sum_{l=1}^{N} \left[ \omega^l (\theta_i) - \overline{\omega}(\theta_i) \right] \left[ \omega^l (\theta_j) - \overline{\omega}(\theta_j) \right]
\end{equation}
In this equation, the summation is over $ N $ independent realizations. However, we used results obtained from a bootstrap resampling analysis (e.g.,~\citealt{Ling_1986, Benoist_1996}) extrapolated from the same parent sample, so they are not independent. When the covariance matrix is estimated from the data itself, a correction factor of $ (N - 1)^2/N $ must be taken into account, and the covariance matrix becomes:
\begin{equation}
C_{ij} = \frac{N_{\text{boot}} - 1}{N_{\text{boot}}} \sum_{l=1}^{N_{\text{boot}}} \left[ \omega^l (\theta_i) - \overline{\omega}(\theta_i) \right] \left[ \omega^l (\theta_j) - \overline{\omega}(\theta_j) \right]
\end{equation}
where $N_{\text{boot}}$ is the total number of resamplings performed, $ \omega^l (\theta_i) $ is the measured cross-correlation function from each realization in the $i$-th bin, and $ \overline{\omega}(\theta_i) $ is the mean of the cross-correlation function in the same bin. We considered the associated uncertainty in each bin to be the square root of the corresponding diagonal element in the covariance matrix.\\
\indent To derive the galaxy bias, we approximate the real-space correlation function using a power-law expression $\xi(r)=(r/r_{0})^{-\gamma}$, with $\gamma$ defined as $\gamma\equiv\beta+1$, i.e., we fix $\gamma$ to 1.6. We derive $r_0$ using the amplitude $A_{w}$ through the Limber transform \citep{Adelberger_2005}:
\begin{equation}\label{eq: correlation length}
  A_{w}=\frac{r_{0}^{\gamma}B[1/2,1/2(\gamma -1)]\int_{0}^{\infty}dzN(z)^{2}f^{1-\gamma}g(z)^{-1}}{[\int_{0}^{\infty}dzN(z)]^{2}}
\end{equation}
In this context, $f$ stands for the transverse comoving distance, given by $f\equiv(1 + z)D_{A}(\theta)$, where $D_A$ is the angular diameter distance. $N(z)$ represents the redshift distribution of dropouts, factoring in source identification efficiency and completeness. The term $B(t_1,t_2)= \Gamma(t_1)\Gamma(t_2)/\Gamma(t_1+t_2)$ denotes the beta function, $g(z)\equiv c/H(z)$ is the comoving distance and $r_0$ represent the correlation length measured in $h^{-1}Mpc$.\\
\indent From the real-space correlation function $\xi(r)$, we define the galaxy bias $b(z)$ as the ratio between the galaxy variance $\sigma_{8,g}$ and the linear matter fluctuation $\sigma_{8}(z)$, both within a sphere of comoving radius $8h^{-1}Mpc$ \citep{Peebles_1980}:
\begin{equation}\label{eq: bias}
  b(z)=\frac{\sigma_{8,g}}{\sigma_{8}(z)}
\end{equation}
assuming $\sigma_{8}(0)=0.828$ from the cosmology model and the galaxy variance expressed as:
\begin{equation}
  \sigma_{8,g}^{2}=\frac{72(r_0/8h^{-1}Mpc)^{\gamma}}{(3-\gamma)(4-\gamma)(6-\gamma)2^{\gamma}}
\end{equation}

\subsection{Random points catalog}\label{subsec: random points}

In Eq.\ref{eq: estimator}, $RR(\theta)$ is the pair count of uniform random points (galaxies) generated in a representative and realistic way that emulates any selection bias of the parent sample. To generate the random points catalog we followed the recovery procedure explained in detail in \citet{Dalmasso_2023}. In short, we used the injection-recovery code \texttt{GLACiAR2} \citep{leethochawalit2022quantitative} to inject galaxies in bins of redshifts ranging from $z_{min}=5.0$ to $z_{max}=11.0$ with an increment of 0.5 and UV magnitudes ranging from $m_{min}=-24.4$ to $m_{max}=32.$ with an increment of 0.5 mag into the JADES images. We injected $N=3200$ galaxies in each redshift-magnitude bin at random positions (roughly one galaxies per 30 arcsec$^2$). The galaxies have disk-type light profiles with Sersic index $n=1$ and random inclinations and ellipticities. The SEDs of the injected galaxies are randomly drawn from the JAGUAR mock catalogue \citep{Williams2018} of the same redshift bin. The galaxies are recovered in a similar fashion to the real data as described in \citet{Rieke_2023} where the detection image is the signal-to-noise image created from F227W, F335M, F56W, F410M and F444W images. The final random catalogues are those that are recovered to be either isolated or minimally overlapping with existing objects.\\
\indent As the random catalog of synthetic sources was generated based on the assumption of a flat input luminosity function $\Phi(m)$, we carried out a Monte Carlo hit-and-miss procedure to establish the final set of random points for our clustering analysis. The procedure consisted of assigning to each element in the random catalog of galaxy a number between zero and one and compare that with the probability of finding that galaxy in an observation. The probability function governing the selection process was formulated as follows:
\begin{equation}
p(m) = \frac{\Phi(m)}{\Phi(m_{\text{lim}})}
\end{equation}
where m is the apparent magnitude of the considered galaxy and $m_{lim}$ is the limit apparent magnitude considered of $m_{F200W}=31.1$ (see Sec.\ref{sec: data selection}). We used the Schechter luminosity function parameters evolution with redshift from \citet{Bouwens_2021}.

\section{Results and Discussion}\label{sec: clustering results}

\begin{table*}
\centering
\begin{tabular}{cccccccccc}
  \hline \hline
  \multicolumn{9}{c}{GOODS-S Full field}\\
  \hline
  $z-bin$ & $\overline{z}$ & $N_{g,tot}$ & $N_{g,sel}$ & $A_{w}$ & $r_{0}/(\mathrm{h^{-1}Mpc})$ & $b$ & $log(M_h/M_{\odot})$ & $\epsilon_{DC}$ & $\chi^2$\\
  (1) & (2) & (3) & (4) & (5) & (6) & (7) & (8) & (9) & (10)\\
  \hline \hline
  $5\leq z<6.0$ & $5.5$ & $3449$ & $2075$ & $0.6\pm0.3$ & $1.5\pm0.5$ & $2.1\pm0.6$ & $9.5^{+0.2}_{-0.2}$ & $0.006^{+0.005}_{-0.003}$ & $0.6$\\[.2cm]
  $6\leq z<7.0$ & $6.5$ & $3880$ & $2786$ & $0.8\pm0.3$ & $2.2\pm0.6$ & $3.1\pm0.5$ & $9.9^{+0.3}_{-0.4}$ & $0.011^{+0.023}_{-0.008}$ & $0.1$\\[.2cm]
  $7\leq z<8.0$ & $7.4$ & $1472$ & $1041$ & $0.9\pm0.4$ & $3.1\pm0.7$ & $4.1\pm0.8$ & $10.1^{+0.4}_{-0.5}$ & $0.012^{+0.040}_{-0.010}$ & $0.8$\\[.2cm]
  $8\leq z<9.0$ & $8.5$ & $982$ & $631$ & $1.8\pm0.4$ & $3.5\pm0.7$ & $5.2\pm0.7$ & $10.15^{+0.3}_{-0.3}$ & $0.012^{+0.032}_{-0.009}$ & $0.9$\\[.2cm]
  $9\leq z<10.0$ & $9.3$ & $426$ & $163$ & $1.9\pm0.6$ & $4.8\pm1.3$ & $7.2\pm1.5$ & $10.47^{+0.3}_{-0.4}$ & $0.045^{+0.230}_{-0.039}$ & $0.7$\\[.2cm]
  $10\leq z<11.0$ & $10.6$ & $207$ & $55$ & $2.5\pm1.1$ & $6.8\pm1.4$ & $9.6\pm1.7$ & $10.60^{+0.3}_{-0.4}$ & $0.075^{+0.540}_{-0.067}$ & $0.2$\\
  \hline \hline
\end{tabular}
\caption{Galaxy clustering parameters obtained from ACF analysis with power-law model with a fixed $\beta=0.6$. (1) Redshift bin. (2) Mean redshift. (3) Number of galaxies in the redshift bin. (4) Number of galaxies selected according to the criteria (Sec.\ref{sec: data selection}). (5) Best-fit value of the amplitude $A_w$. (6) Best-fit value of the correlation length $r_0$ in units of $\mathrm{h^{-1}Mpc}$. (7) Galaxy bias. (8) Dark matter halo mass (log scale) in units of $M_{\odot}$. (9) Duty cycle. (10) Reduced $\chi^2$ value.}
\label{tab: clustering results}
\end{table*}

\subsection{Angular correlation function and Galaxy bias}\label{subsec: acf and bias}

\begin{figure*}
  \includegraphics[width=1\linewidth]{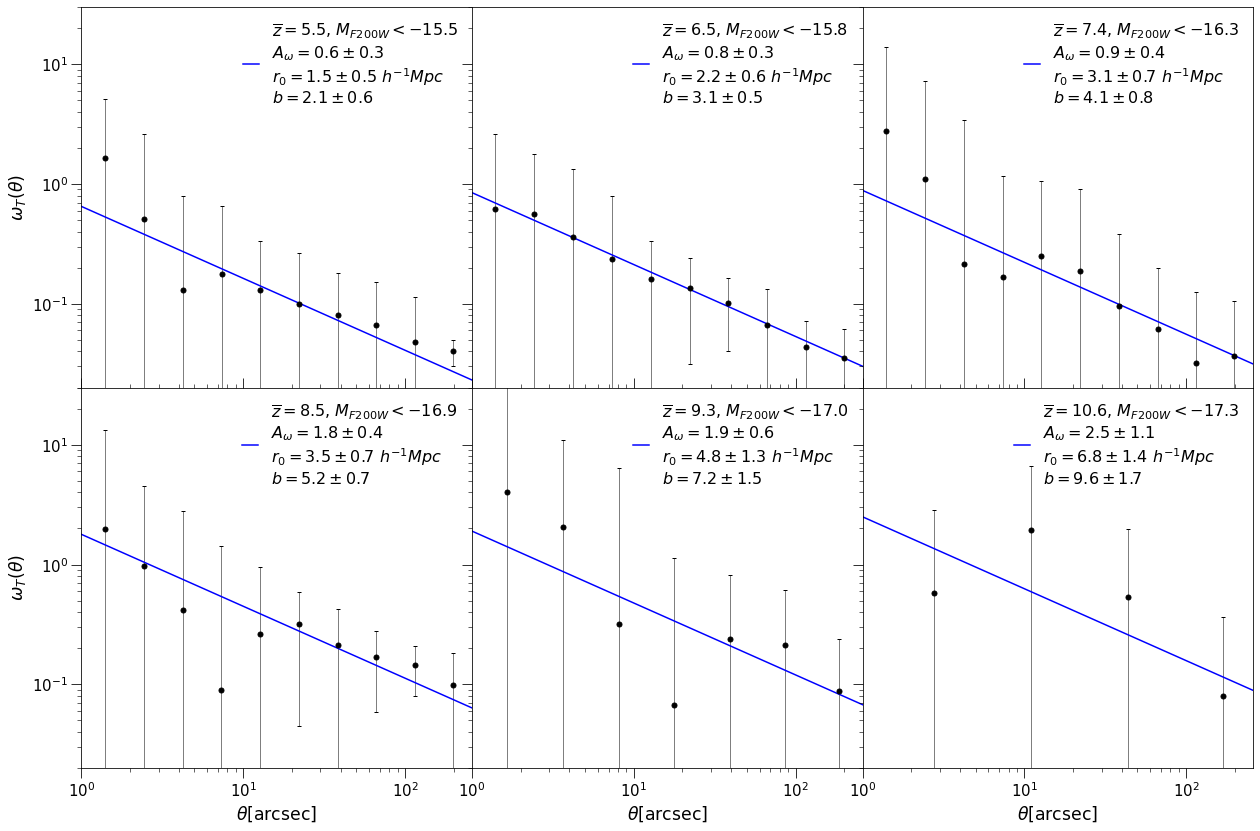}
\caption{Measured ACFs in the GOODS-S field at six different mean redshifts. The black dots denote the ACF measurements with associated $1\sigma$ uncertainties, the blue solid lines represent a the power law best-fit with parameters listed in the legend box.}
\label{fig: Angular correlation function}
\end{figure*}

\begin{figure}
\includegraphics[width=1.0\linewidth]{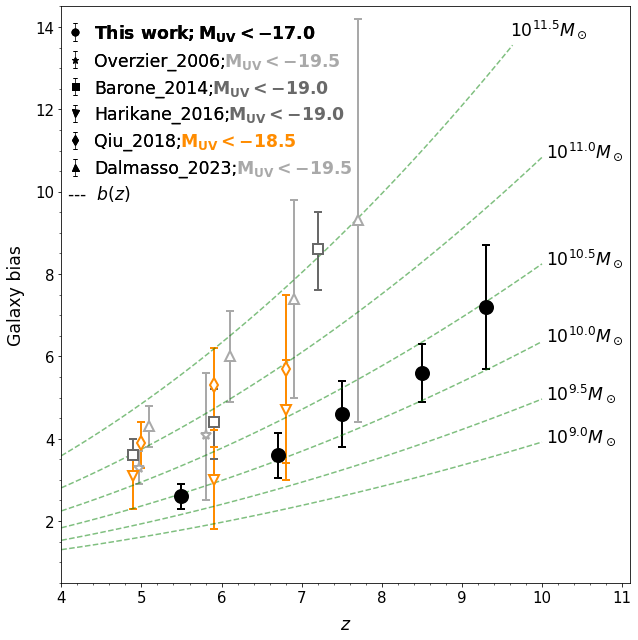}
\caption{Galaxy bias evolution as a function of redshift $z$. Circles represent the results obtained in this work. Other symbols are previous bias measurements: stars \citet{Overzier_2006}, squares \citet{Barone_2014}, downward triangles \citet{Harikane_2016}, diamonds \citet{Qiu_2018} and upward triangles \citet{Dalmasso_2023}. Green dashed lines represent the theoretical bias evolution at a fixed dark matter halo mass $M_h$ $[M_{\odot}]$. Note that our observations are substantially deeper than other studies, hence the inferred bias is lower.}
\label{fig: bias vs z}
\end{figure}

Fig.\ref{fig: Angular correlation function} presents the ACF estimated using LBG candidates from the GOODS-S field. The lower right panel features the newly obtained measurement at redshift $z=10.6$. Also, within the same figure we present results spanning the redshift range $5.0\leq z<10.0$ to show the evolution of the ACF with redshift and to compare our new measurements from JWST data with those reported in the literature and based on other facilities. Each panel also reports two clustering parameters: $A_w$, the coefficient obtained from fitting the angular correlation function using Eq.\ref{eq: estimator model}, and the correlation length $r_0$ derived from the fit (both with respective uncertainties at 1$\sigma$). The parameters obtained are tabulated in columns (5) and (6) in Tab.\ref{tab: clustering results}.\\
\indent With a simple functional model (Eq.\ref{eq: estimator model}) to fit our measurements, we get overall agreement with the data, even at high redshifts. Interestingly, the measurement and fit remains of acceptable quality even when number of sources is low, as exemplified by the $N_g=55$ candidates at redshift $z=10.6$ (bottom right panel of Fig.\ref{fig: Angular correlation function}). This determination at redshift $z=10.6$ marks the first measurement of galaxy clustering at less than 500 Myr since the Big Bang based on sources substantially fainter than previously reported with $M_{F200W}\lesssim-17.3$. The galaxy bias measurement can also be used to determine the characteristic dark matter halo mass that hosts the observed galaxies. For a fixed redshift, this is done by computing the theoretical dark matter halo bias $b(M_h)$ as a function of halo mass $M_h$ as described by \citet{Tormen_1999} and by numerically inverting the relationship to derive $M_h(b)$. Using the results presented in Fig.\ref{fig: Angular correlation function} and reported in column (6) in Tab.\ref{tab: clustering results} we obtain the corresponding dark matter halo masses in column (8). Our results confirm the trends in redshift evolution of clustering reported in the literature, specifically the correlation length and bias steadily increase with redshift. In Tab.\ref{tab: clustering results}, the last column displays the values of the reduced $\chi^2$ to offer information on the single power law fit quality.\\
\indent In Fig.\ref{fig: bias vs z}, the evolution of galaxy bias is presented for ACF measurements obtained with galaxy candidates brighter than $M_{F200W}\lesssim-17.0$ at $z<10$. Fig.~\ref{fig: bias vs z} also reports bias measurements in the literature, based primarily on Hubble Space Telescope and/or Subaru observations (\citealt{Overzier_2006,Barone_2014,Harikane_2016,Qiu_2018,Dalmasso_2023}). It should be noted that a direct comparison is not possible, as the JWST observations are substantially deeper. However, there is qualitative consistency since the bias we measure is lower than the previous measurements for brighter sources at fixed redshift. Intrinsically, our observations show a clear trend of increasing bias with redshift and a slight increase characteristic halo mass as well, which we interpret as a result of the evolution of the luminosity distance, which makes a magnitude limited survey probe brighter (and therefore more massive objects) at increasing redshift (e.g.,~\citealt{Mason_2015,Trenti_2015}).\\

\subsection{Galaxy bias magnitude dependence}\label{subsec: magnitude bias dependence}

\begin{figure}
\includegraphics[width=1.0\linewidth]{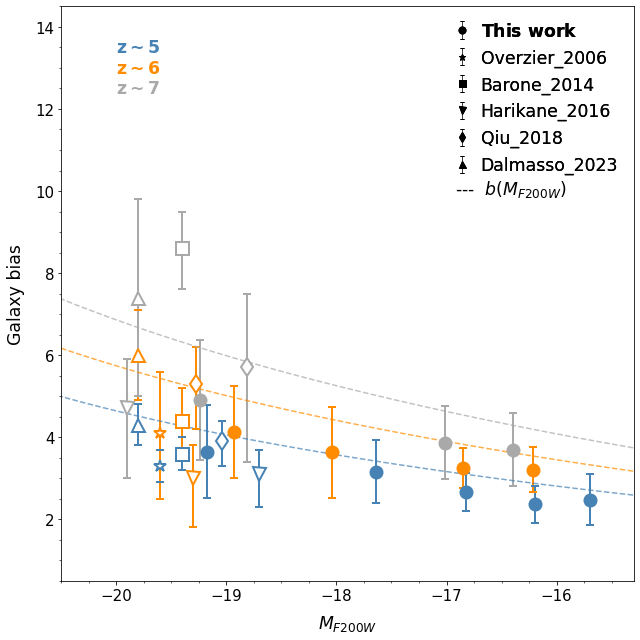}
\caption{Galaxy bias evolution as a function of absolute magnitude $M_{F200W}$ for three different redshift bins (blue, orange and grey for $z=5,6,7$ respectively. Circles represent the results obtained in this work. Other symbols are literature measurements: stars \citet{Overzier_2006}, squares \citet{Barone_2014}, downward triangles \citet{Harikane_2016}, diamonds \citet{Qiu_2018} and upward triangles \citet{Dalmasso_2023}. Colored dashed lines represent the theoretical bias evolution model as a function of absolute magnitude $M_{F200W}$, derived from \citet{Trenti_2015}.}
\label{fig: bias vs M}
\end{figure}

Interestingly, our sample is sufficiently large to explore luminosity dependence on galaxy bias by building subsamples of sources selected at different magnitude intervals for sources at redshifts $z\sim5$, $z\sim6$, and $z\sim7$ (sources at higher redshift are too few to enable a split in magnitude subsamples). The magnitude bin size for each redshift selection has been chosen to ensure a balanced distribution of sources in each subsample, resulting in the average $M_{F200W}$ for each bin shown in Fig.\ref{fig: bias vs M}.\\
\indent For comparison in Fig.\ref{fig: bias vs M} we also report results from prior studies on galaxy clustering. Reassuringly, with a consistent luminosity selection, our bias determination for the brightest samples are overlapping with prior measurements reported in the literature (\citealt{Overzier_2006,Barone_2014,Harikane_2016,Qiu_2018,Dalmasso_2023}). This agreement gives both confidence in our analysis, and also confirms that previous photometric samples of candidate galaxies at the end of the epoch of reionization have the same clustering properties as the higher quality JADES samples (which benefit from deeper data for improved photometric selection).\\
\indent Fig.\ref{fig: bias vs M} shows a decrease of bias with decreasing galaxy luminosity out to $M_{F200W}\sim -15.5$. This observed trend is effectively captured by the theoretical model for the bias evolution $b(M_{F200W})$ derived from \citet{Trenti_2015} using the Gamma Ray Burst no-metal bias scenario, which traces star formation and therefore gives a halo mass-UV magnitude relationship we can plot against our data.\\
\indent As an additional check for the robustness of estimated errors, we considered a larger catalog of random points $(N_{r}^{'} = 20 \cdot N_r)$, finding that the results are consistent with each other, with a 2\% deviation from results obtained using a smaller random catalog.\\
\indent In passing, we note that if we were to extend our analysis to fainter sources in the magnitude regime around $M_{F200W} \sim -14.0$, we would derive a galaxy bias of $b \simeq 7$ at redshift $z=5$, i.e. a significant increase over the decreasing trend presented in Fig.~\ref{fig: bias vs M}. Such high value would be inconsistent with theoretical predictions for clustering of dwarf galaxies. We attribute the high measurement to spatially-dependent systematic error estimation of completeness for the faintest sources detectable by JWST that are not fully captured by our source recovery simulations because of the extreme faintness of such sources, which results in very low completeness. This provides supports for our choice to limit the analysis to sources with brightness $m_{F200W} < 31.1$ ($M_{F200W} < -15.5$ at $z\simeq5.5$) in order to achieve robust clustering measurements.

\subsection{Abundance matching and Duty Cycle}\label{subsec: duty cycle}

To explore the connection between dark mass and star formation at high redshift, we can resort to abundance matching and duty cycle modelling. This process entails statistically correlating the mass of dark matter halos with the luminosity (or stellar mass) of galaxies in a given sample. This methodology is based on the premise that the most luminous galaxy inhabits the most massive halo, followed by the second most massive galaxy residing in the subsequent most massive halo, and so forth. By doing so, it is possible to infer the intricate relationship between galaxy formation and the underlying dark matter structure (but see \citealt{Ren_2019}). This relationship serves as the basis for calculating the duty cycle $\epsilon_{DC}$, which is defined as the fraction of time a dark matter halo of a certain mass hosts a UV-bright LBGs (e.g.,\citealt{martinez_2002,Vale_2004}). The fundamental equation for conducting abundance matching is:
\begin{equation}\label{eq:DC}
  \epsilon_{DC}\int_{M_h}^{+\infty}n(M_h,z)dM_h = \int_{L_g}^{+\infty}\Phi(L,z)dL
\end{equation}
Here, $\Phi(L,z)$ represents the Luminosity Function (LF) at redshift $z$, $n(M_h,z)$ stands for the Dark Matter Halo Mass Function (MF) \citep{Tormen_1999} and ($M_h$,$L_g$) are respectively the the minimum dark matter halo mass and the minimum luminosity of the galaxy. At a fixed value of $L_g$, Eq.\ref{eq:DC} leads to the determination of a one-to-one relationship between $\epsilon_{DC}$ and $M_{h}$ that satisfies the equation constraint. The degeneracy between the two quantities can be resolved through clustering (bias) measurements, which independently define the relationship between $L_g$ and $M_h$. This is because bias increases with halo mass and the luminosity of galaxies, as more massive halos tend to host more luminous galaxies, thus affecting the overall distribution of halos \citep{Tormen_1999}. When it comes to the luminosity function parameters, we employed the parameters that exhibit evolution with redshift, as presented in \citet{Bouwens_2021}. The obtained duty cycle values are presented in the last column of Tab.\ref{tab: clustering results}.\\
\indent In our findings, we observe a growing trend linked to the redshift of the sample, peaking at a maximum duty cycle value of $\epsilon_{DC}=0.075^{+0.540}_{-0.067}$. This observation, observed at our highest redshift of $z=10.6$, aligns with the theoretical predictions outlined in \citealt{wyithe_2014}. The authors emphasized that for $z > 4$, a low duty cycle ($\epsilon_{DC}\lesssim0.15$) is necessary to reconcile the observed star formation rate with the relatively low observed total stellar mass.
Within the redshift range examined in our study, we find duty cycle values below this theoretical threshold, confirming the existence of a previously unidentified population of UV-faint galaxies for $z>4$.

\section{Summary and perspectives}\label{sec: summary}

In this paper, we utilised the methodology outlined and discussed in \citet{Dalmasso_2023} to carry out galaxy clustering measurements well into the epoch of reionization using faint sources ($M_{F200W}$<-17.3 at z=10.6) using JWST data from the guaranteed time observation program JWST Advanced Deep Extragalactic Survey (JADES)(\citealt{Bunker_2023,Eisenstein_2023,Hainline_2023,Rieke_2023}) with the NIRCam instrument. Our main results and conclusions are as follows:\\
\indent $\bullet$ We derived a galaxy bias measurement of $b = 9.6\pm1.7$ from a sample of $N_g = 55$ candidates at redshift $z = 10.6$ with a cut in absolute magnitude of $M_{F200W}<-17.3$.\\
\indent $\bullet$ We studies self-consistently the clustering evolution from $5.0\leq z<10$. 
This allowed us to delineate the evolution of galaxy bias, revealing an increasing trend as anticipated by theory. Specifically, we report a value of $b=2.1\pm0.6$ at $z=5.5$ and $b=7.2\pm1.5$ at redshift $z=9.3$ (see Tab.\ref{tab: clustering results}).\\
\indent $\bullet$ We studied the relationship between galaxy bias and magnitude (Fig.\ref{fig: bias vs M}). Our results at the faintest magnitudes are in line with theoretical predictions out to $M_{F200W} \sim -15.5$ at $z\sim5$, and they present an evolution consistent with previous studies carried out for brighter sources.\\
\indent Through this paper, we have demonstrated the viability of conducting clustering measurements on LBG samples observed during the epoch of reionization, even as the candidates become increasingly faint and rare, thanks to the power of JWST infrared instrumentation. As photometric catalogs extend to even fainter magnitudes and the pool of available candidates expands from the growing number of JWST observations, the initial results reported here have the potential to blossom into a quantitative dataset that can be compared to more sophisticated halo occupation models (e.g., see ~\citealt{Harikane_2022}) to understand how dark matter halos transform gas into stars the earliest times in the Universe history. 

\section*{Acknowledgements}
We thank the anonymous referee for useful suggestions and
comments that have improved the manuscript. This research was supported in part by the Australian Research Council Centre of Excellence for All Sky Astrophysics in 3 Dimensions (ASTRO 3D), through project number CE170100013. We thank Benjamin Metha for helpful comments and suggestions, and we acknowledge useful suggestions from an anonymous referee.

\section*{Data Availability}

The data used to conduct the analysis are from "The JWST Advanced Deep Extragalactic Survey (JADES)" \citep{Bunker_2023,Eisenstein_2023,Hainline_2023,Rieke_2023}.



\bibliographystyle{mnras}
\bibliography{main} 

\begin{thebibliography}{}
\makeatletter
\relax
\def\mn@urlcharsother{\let\do\@makeother \do\$\do\&\do\#\do\^\do\_\do\%\do\~}
\def\mn@doi{\begingroup\mn@urlcharsother \@ifnextchar [ {\mn@doi@} {\mn@doi@[]}}
\def\mn@doi@[#1]#2{\def\@tempa{#1}\ifx\@tempa\@empty \href {http://dx.doi.org/#2} {doi:#2}\else \href {http://dx.doi.org/#2} {#1}\fi \endgroup}
\def\mn@eprint#1#2{\mn@eprint@#1:#2::\@nil}
\def\mn@eprint@arXiv#1{\href {http://arxiv.org/abs/#1} {{\tt arXiv:#1}}}
\def\mn@eprint@dblp#1{\href {http://dblp.uni-trier.de/rec/bibtex/#1.xml} {dblp:#1}}
\def\mn@eprint@#1:#2:#3:#4\@nil{\def\@tempa {#1}\def\@tempb {#2}\def\@tempc {#3}\ifx \@tempc \@empty \let \@tempc \@tempb \let \@tempb \@tempa \fi \ifx \@tempb \@empty \def\@tempb {arXiv}\fi \@ifundefined {mn@eprint@\@tempb}{\@tempb:\@tempc}{\expandafter \expandafter \csname mn@eprint@\@tempb\endcsname \expandafter{\@tempc}}}

\bibitem[\protect\citeauthoryear{Ade et~al.,}{Ade et~al.}{2014}]{ade2014planck}
Ade P.~A.,  et~al., 2014, Astronomy \& Astrophysics, 571, A16

\bibitem[\protect\citeauthoryear{Adelberger, Steidel, Pettini, Shapley, Reddy  \& Erb}{Adelberger et~al.}{2005}]{Adelberger_2005}
Adelberger K.~L.,  Steidel C.~C.,  Pettini M.,  Shapley A.~E.,  Reddy N.~A.,   Erb D.~K.,  2005, \mn@doi [The Astrophysical Journal] {10.1086/426580}, 619, 697

\bibitem[\protect\citeauthoryear{Akhlaghi \& Ichikawa}{Akhlaghi \& Ichikawa}{2015}]{Akhlaghi_2015}
Akhlaghi M.,  Ichikawa T.,  2015, \mn@doi [The Astrophysical Journal Supplement Series] {10.1088/0067-0049/220/1/1}, 220, 1

\bibitem[\protect\citeauthoryear{{AstropyCollaboration}}{{AstropyCollaboration}}{2022}]{Astropy_Collaboration_2022}
{AstropyCollaboration} 2022, \mn@doi [The Astrophysical Journal] {10.3847/1538-4357/ac7c74}, 935, 167

\bibitem[\protect\citeauthoryear{{Bahcall} \& {Soneira}}{{Bahcall} \& {Soneira}}{1980}]{Bahcall_1983}
{Bahcall} J.~N.,  {Soneira} R.~M.,  1980, \mn@doi [\apjs] {10.1086/190685}, \href {https://ui.adsabs.harvard.edu/abs/1980ApJS...44...73B} {44, 73}

\bibitem[\protect\citeauthoryear{{Bardeen}, {Bond}, {Kaiser}  \& {Szalay}}{{Bardeen} et~al.}{1986}]{Bardeen_1986}
{Bardeen} J.~M.,  {Bond} J.~R.,  {Kaiser} N.,   {Szalay} A.~S.,  1986, \mn@doi [\apj] {10.1086/164143}, \href {https://ui.adsabs.harvard.edu/abs/1986ApJ...304...15B} {304, 15}

\bibitem[\protect\citeauthoryear{Barone-Nugent et~al.,}{Barone-Nugent et~al.}{2014}]{Barone_2014}
Barone-Nugent R.~L.,  et~al., 2014, \mn@doi [The Astrophysical Journal] {10.1088/0004-637x/793/1/17}, 793, 17

\bibitem[\protect\citeauthoryear{{Benoist}, {Maurogordato}, {da Costa}, {Cappi}  \& {Schaeffer}}{{Benoist} et~al.}{1996}]{Benoist_1996}
{Benoist} C.,  {Maurogordato} S.,  {da Costa} L.~N.,  {Cappi} A.,   {Schaeffer} R.,  1996, \mn@doi [\apj] {10.1086/178078}, \href {https://ui.adsabs.harvard.edu/abs/1996ApJ...472..452B} {472, 452}

\bibitem[\protect\citeauthoryear{Bian et~al.,}{Bian et~al.}{2013}]{Bian_2013}
Bian F.,  et~al., 2013, \mn@doi [The Astrophysical Journal] {10.1088/0004-637X/774/1/28}, 774, 28

\bibitem[\protect\citeauthoryear{Bouwens et~al.,}{Bouwens et~al.}{2021}]{Bouwens_2021}
Bouwens R.~J.,  et~al., 2021, \mn@doi [The Astronomical Journal] {10.3847/1538-3881/abf83e}, 162, 47

\bibitem[\protect\citeauthoryear{Bradley et~al.,}{Bradley et~al.}{2023}]{Bradley_2023}
Bradley L.,  et~al., 2023, astropy/photutils: 1.8.0, \mn@doi{10.5281/zenodo.7946442}, \url {https://doi.org/10.5281/zenodo.7946442}

\bibitem[\protect\citeauthoryear{Brammer, van Dokkum  \& Coppi}{Brammer et~al.}{2008}]{Brammer_2008}
Brammer G.~B.,  van Dokkum P.~G.,   Coppi P.,  2008, \mn@doi [The Astrophysical Journal] {10.1086/591786}, 686, 1503

\bibitem[\protect\citeauthoryear{{Bullock}, {Kolatt}, {Sigad}, {Somerville}, {Kravtsov}, {Klypin}, {Primack}  \& {Dekel}}{{Bullock} et~al.}{2001}]{Bullock_2001}
{Bullock} J.~S.,  {Kolatt} T.~S.,  {Sigad} Y.,  {Somerville} R.~S.,  {Kravtsov} A.~V.,  {Klypin} A.~A.,  {Primack} J.~R.,   {Dekel} A.,  2001, \mn@doi [\mnras] {10.1046/j.1365-8711.2001.04068.x}, \href {https://ui.adsabs.harvard.edu/abs/2001MNRAS.321..559B} {321, 559}

\bibitem[\protect\citeauthoryear{{Bunker} et~al.,}{{Bunker} et~al.}{2023}]{Bunker_2023}
{Bunker} A.~J.,  et~al., 2023, \mn@doi [arXiv e-prints] {10.48550/arXiv.2306.02467}, \href {https://ui.adsabs.harvard.edu/abs/2023arXiv230602467B} {p. arXiv:2306.02467}

\bibitem[\protect\citeauthoryear{Dalmasso, Trenti  \& Leethochawalit}{Dalmasso et~al.}{2023}]{Dalmasso_2023}
Dalmasso N.,  Trenti M.,   Leethochawalit N.,  2023, \mn@doi [Monthly Notices of the Royal Astronomical Society] {10.1093/mnras/stad3901}, p. stad3901

\bibitem[\protect\citeauthoryear{{Davis} \& {Peebles}}{{Davis} \& {Peebles}}{1983}]{Davis_1983}
{Davis} M.,  {Peebles} P.~J.~E.,  1983, \mn@doi [\apj] {10.1086/160884}, \href {https://ui.adsabs.harvard.edu/abs/1983ApJ...267..465D} {267, 465}

\bibitem[\protect\citeauthoryear{{Eisenstein} et~al.,}{{Eisenstein} et~al.}{2023}]{Eisenstein_2023}
{Eisenstein} D.~J.,  et~al., 2023, \mn@doi [arXiv e-prints] {10.48550/arXiv.2306.02465}, \href {https://ui.adsabs.harvard.edu/abs/2023arXiv230602465E} {p. arXiv:2306.02465}

\bibitem[\protect\citeauthoryear{{Foucaud}, {McCracken}, {Le F{\`e}vre}, {Arnouts}, {Brodwin}, {Lilly}, {Crampton}  \& {Mellier}}{{Foucaud} et~al.}{2003}]{Foucaud_2003}
{Foucaud} S.,  {McCracken} H.~J.,  {Le F{\`e}vre} O.,  {Arnouts} S.,  {Brodwin} M.,  {Lilly} S.~J.,  {Crampton} D.,   {Mellier} Y.,  2003, \mn@doi [\aap] {10.1051/0004-6361:20031181}, \href {https://ui.adsabs.harvard.edu/abs/2003A&A...409..835F} {409, 835}

\bibitem[\protect\citeauthoryear{Giavalisco \& Dickinson}{Giavalisco \& Dickinson}{2001}]{Giavalisco_2001}
Giavalisco M.,  Dickinson M.,  2001, \mn@doi [The Astrophysical Journal] {10.1086/319715}, 550, 177

\bibitem[\protect\citeauthoryear{{Hainline} et~al.,}{{Hainline} et~al.}{2023}]{Hainline_2023}
{Hainline} K.~N.,  et~al., 2023, \mn@doi [arXiv e-prints] {10.48550/arXiv.2306.02468}, \href {https://ui.adsabs.harvard.edu/abs/2023arXiv230602468H} {p. arXiv:2306.02468}

\bibitem[\protect\citeauthoryear{Harikane et~al.,}{Harikane et~al.}{2016}]{Harikane_2016}
Harikane Y.,  et~al., 2016, \mn@doi [The Astrophysical Journal] {10.3847/0004-637X/821/2/123}, 821, 123

\bibitem[\protect\citeauthoryear{Harikane et~al.,}{Harikane et~al.}{2022}]{Harikane_2022}
Harikane Y.,  et~al., 2022, \mn@doi [The Astrophysical Journal Supplement Series] {10.3847/1538-4365/ac3dfc}, 259, 20

\bibitem[\protect\citeauthoryear{Hatfield, Bowler, Jarvis  \& Hale}{Hatfield et~al.}{2018}]{Hatfield_2018}
Hatfield P.~W.,  Bowler R. A.~A.,  Jarvis M.~J.,   Hale C.~L.,  2018, \mn@doi [Monthly Notices of the Royal Astronomical Society] {10.1093/mnras/sty856}, 477, 3760

\bibitem[\protect\citeauthoryear{Jenkins, Frenk, White, Colberg, Cole, Evrard, Couchman  \& Yoshida}{Jenkins et~al.}{2001}]{Jenkins_2001}
Jenkins A.,  Frenk C.,  White S.~D.,  Colberg J.~M.,  Cole S.,  Evrard A.~E.,  Couchman H.,   Yoshida N.,  2001, Monthly Notices of the Royal Astronomical Society, 321, 372

\bibitem[\protect\citeauthoryear{Klypin, Primack  \& Holtzman}{Klypin et~al.}{1996}]{Klypin_1996}
Klypin A.,  Primack J.~R.,   Holtzman J.~A.,  1996, The Astrophysical Journal, 466, 13

\bibitem[\protect\citeauthoryear{{Landy} \& {Szalay}}{{Landy} \& {Szalay}}{1993}]{1993ApJ...412...64L}
{Landy} S.~D.,  {Szalay} A.~S.,  1993, \mn@doi [\apj] {10.1086/172900}, \href {https://ui.adsabs.harvard.edu/abs/1993ApJ...412...64L} {412, 64}

\bibitem[\protect\citeauthoryear{Lee, Giavalisco, Gnedin, Somerville, Ferguson, Dickinson  \& Ouchi}{Lee et~al.}{2006}]{Lee_2006}
Lee K.,  Giavalisco M.,  Gnedin O.~Y.,  Somerville R.~S.,  Ferguson H.~C.,  Dickinson M.,   Ouchi M.,  2006, \mn@doi [The Astrophysical Journal] {10.1086/500387}, 642, 63

\bibitem[\protect\citeauthoryear{Leethochawalit, Trenti, Morishita, Roberts-Borsani  \& Treu}{Leethochawalit et~al.}{2022}]{leethochawalit2022quantitative}
Leethochawalit N.,  Trenti M.,  Morishita T.,  Roberts-Borsani G.,   Treu T.,  2022, Monthly Notices of the Royal Astronomical Society, 509, 5836

\bibitem[\protect\citeauthoryear{Ling, Frenk  \& Barrow}{Ling et~al.}{1986}]{Ling_1986}
Ling E.~N.,  Frenk C.~S.,   Barrow J.~D.,  1986, \mn@doi [Monthly Notices of the Royal Astronomical Society] {10.1093/mnras/223.1.21P}, 223, 21P

\bibitem[\protect\citeauthoryear{{Mart{\'\i}nez}, {Zandivarez}, {Merch{\'a}n}  \& {Dom{\'\i}nguez}}{{Mart{\'\i}nez} et~al.}{2002}]{martinez_2002}
{Mart{\'\i}nez} H.~J.,  {Zandivarez} A.,  {Merch{\'a}n} M.~E.,   {Dom{\'\i}nguez} M.~J.~L.,  2002, \mn@doi [\mnras] {10.1046/j.1365-8711.2002.06020.x}, \href {https://ui.adsabs.harvard.edu/abs/2002MNRAS.337.1441M} {337, 1441}

\bibitem[\protect\citeauthoryear{Mason, Trenti  \& Treu}{Mason et~al.}{2015}]{Mason_2015}
Mason C.~A.,  Trenti M.,   Treu T.,  2015, \mn@doi [The Astrophysical Journal] {10.1088/0004-637X/813/1/21}, 813, 21

\bibitem[\protect\citeauthoryear{{Mo} \& {White}}{{Mo} \& {White}}{1996}]{Mo_1996}
{Mo} H.~J.,  {White} S.~D.~M.,  1996, \mn@doi [\mnras] {10.1093/mnras/282.2.347}, \href {https://ui.adsabs.harvard.edu/abs/1996MNRAS.282..347M} {282, 347}

\bibitem[\protect\citeauthoryear{{Oke} \& {Gunn}}{{Oke} \& {Gunn}}{1983}]{1983ApJ...266..713O}
{Oke} J.~B.,  {Gunn} J.~E.,  1983, \mn@doi [\apj] {10.1086/160817}, \href {https://ui.adsabs.harvard.edu/abs/1983ApJ...266..713O} {266, 713}

\bibitem[\protect\citeauthoryear{Okuta, Unno, Nishino, Hido  \& Loomis}{Okuta et~al.}{2017}]{cupy_2017}
Okuta R.,  Unno Y.,  Nishino D.,  Hido S.,   Loomis C.,  2017, in Proceedings of Workshop on Machine Learning Systems (LearningSys) in The Thirty-first Annual Conference on Neural Information Processing Systems (NIPS). \url {http://learningsys.org/nips17/assets/papers/paper_16.pdf}

\bibitem[\protect\citeauthoryear{Ouchi et~al.,}{Ouchi et~al.}{2004}]{Ouchi_2004}
Ouchi M.,  et~al., 2004, \mn@doi [The Astrophysical Journal] {10.1086/422208}, 611, 685

\bibitem[\protect\citeauthoryear{Ouchi et~al.,}{Ouchi et~al.}{2005}]{Ouchi_2005}
Ouchi M.,  et~al., 2005, \mn@doi [The Astrophysical Journal] {10.1086/499519}, 635, L117

\bibitem[\protect\citeauthoryear{Overzier, Bouwens, Illingworth  \& Franx}{Overzier et~al.}{2006}]{Overzier_2006}
Overzier R.~A.,  Bouwens R.~J.,  Illingworth G.~D.,   Franx M.,  2006, \mn@doi [The Astrophysical Journal] {10.1086/507678}, 648, l5

\bibitem[\protect\citeauthoryear{Peacock \& Nicholson}{Peacock \& Nicholson}{1991}]{10.1093/mnras/253.2.307}
Peacock J.~A.,  Nicholson D.,  1991, \mn@doi [Monthly Notices of the Royal Astronomical Society] {10.1093/mnras/253.2.307}, 253, 307

\bibitem[\protect\citeauthoryear{{Peebles}}{{Peebles}}{1980}]{Peebles_1980}
{Peebles} P.~J.~E.,  1980, {The large-scale structure of the universe}

\bibitem[\protect\citeauthoryear{{Persic} \& {Salucci}}{{Persic} \& {Salucci}}{1992}]{Persic_1992}
{Persic} M.,  {Salucci} P.,  1992, \mn@doi [\mnras] {10.1093/mnras/258.1.14P}, \href {https://ui.adsabs.harvard.edu/abs/1992MNRAS.258P..14P} {258, 14P}

\bibitem[\protect\citeauthoryear{{Qiu} et~al.,}{{Qiu} et~al.}{2018}]{Qiu_2018}
{Qiu} Y.,  et~al., 2018, \mn@doi [\mnras] {10.1093/mnras/sty2633}, \href {https://ui.adsabs.harvard.edu/abs/2018MNRAS.481.4885Q} {481, 4885}

\bibitem[\protect\citeauthoryear{{Ren}, {Trenti}  \& {Mason}}{{Ren} et~al.}{2019}]{Ren_2019}
{Ren} K.,  {Trenti} M.,   {Mason} C.~A.,  2019, \mn@doi [\apj] {10.3847/1538-4357/ab2117}, \href {https://ui.adsabs.harvard.edu/abs/2019ApJ...878..114R} {878, 114}

\bibitem[\protect\citeauthoryear{{Rieke} et~al.,}{{Rieke} et~al.}{2023}]{Rieke_2023}
{Rieke} M.~J.,  et~al., 2023, \mn@doi [arXiv e-prints] {10.48550/arXiv.2306.02466}, \href {https://ui.adsabs.harvard.edu/abs/2023arXiv230602466R} {p. arXiv:2306.02466}

\bibitem[\protect\citeauthoryear{Savoy, Sawicki, Thompson  \& Sato}{Savoy et~al.}{2011}]{Savoy_2011}
Savoy J.,  Sawicki M.,  Thompson D.,   Sato T.,  2011, \mn@doi [The Astrophysical Journal] {10.1088/0004-637X/737/2/92}, 737, 92

\bibitem[\protect\citeauthoryear{Sheth \& Tormen}{Sheth \& Tormen}{1999}]{Tormen_1999}
Sheth R.~K.,  Tormen G.,  1999, \mn@doi [Monthly Notices of the Royal Astronomical Society] {10.1046/j.1365-8711.1999.02692.x}, 308, 119

\bibitem[\protect\citeauthoryear{{Tinker}, {Robertson}, {Kravtsov}, {Klypin}, {Warren}, {Yepes}  \& {Gottl{\"o}ber}}{{Tinker} et~al.}{2010}]{Tinker_2010}
{Tinker} J.~L.,  {Robertson} B.~E.,  {Kravtsov} A.~V.,  {Klypin} A.,  {Warren} M.~S.,  {Yepes} G.,   {Gottl{\"o}ber} S.,  2010, \mn@doi [\apj] {10.1088/0004-637X/724/2/878}, \href {https://ui.adsabs.harvard.edu/abs/2010ApJ...724..878T} {724, 878}

\bibitem[\protect\citeauthoryear{{Trenti}, {Stiavelli}, {Bouwens}, {Oesch}, {Shull}, {Illingworth}, {Bradley}  \& {Carollo}}{{Trenti} et~al.}{2010}]{Trenti_2010}
{Trenti} M.,  {Stiavelli} M.,  {Bouwens} R.~J.,  {Oesch} P.,  {Shull} J.~M.,  {Illingworth} G.~D.,  {Bradley} L.~D.,   {Carollo} C.~M.,  2010, \mn@doi [\apjl] {10.1088/2041-8205/714/2/L202}, \href {https://ui.adsabs.harvard.edu/abs/2010ApJ...714L.202T} {714, L202}

\bibitem[\protect\citeauthoryear{{Trenti}, {Perna}  \& {Jimenez}}{{Trenti} et~al.}{2015}]{Trenti_2015}
{Trenti} M.,  {Perna} R.,   {Jimenez} R.,  2015, \mn@doi [\apj] {10.1088/0004-637X/802/2/103}, \href {https://ui.adsabs.harvard.edu/abs/2015ApJ...802..103T} {802, 103}

\bibitem[\protect\citeauthoryear{Vale \& Ostriker}{Vale \& Ostriker}{2004}]{Vale_2004}
Vale A.,  Ostriker J.~P.,  2004, \mn@doi [Monthly Notices of the Royal Astronomical Society] {10.1111/j.1365-2966.2004.08059.x}, 353, 189

\bibitem[\protect\citeauthoryear{{Williams} et~al.,}{{Williams} et~al.}{2018}]{Williams2018}
{Williams} C.~C.,  et~al., 2018, \mn@doi [\apjs] {10.3847/1538-4365/aabcbb}, \href {https://ui.adsabs.harvard.edu/abs/2018ApJS..236...33W} {236, 33}

\bibitem[\protect\citeauthoryear{Wyithe, Loeb  \& Oesch}{Wyithe et~al.}{2014}]{wyithe_2014}
Wyithe J. S.~B.,  Loeb A.,   Oesch P.~A.,  2014, \mn@doi [Monthly Notices of the Royal Astronomical Society] {10.1093/mnras/stu038}, 439, 1326

\bibitem[\protect\citeauthoryear{{van der Walt} et~al.,}{{van der Walt} et~al.}{2014}]{van_der_Walt_2014}
{van der Walt} S.,  et~al., 2014, \mn@doi [PeerJ] {10.7717/peerj.453}, \href {https://ui.adsabs.harvard.edu/abs/2014PeerJ...2..453V} {2, e453}

\makeatother
\end{thebibliography}



\appendix
\section{ACF measured for the bias evolution with magnitude}\label{app: ACF multimag}

\begin{figure*}
\includegraphics[width=1.0\linewidth]{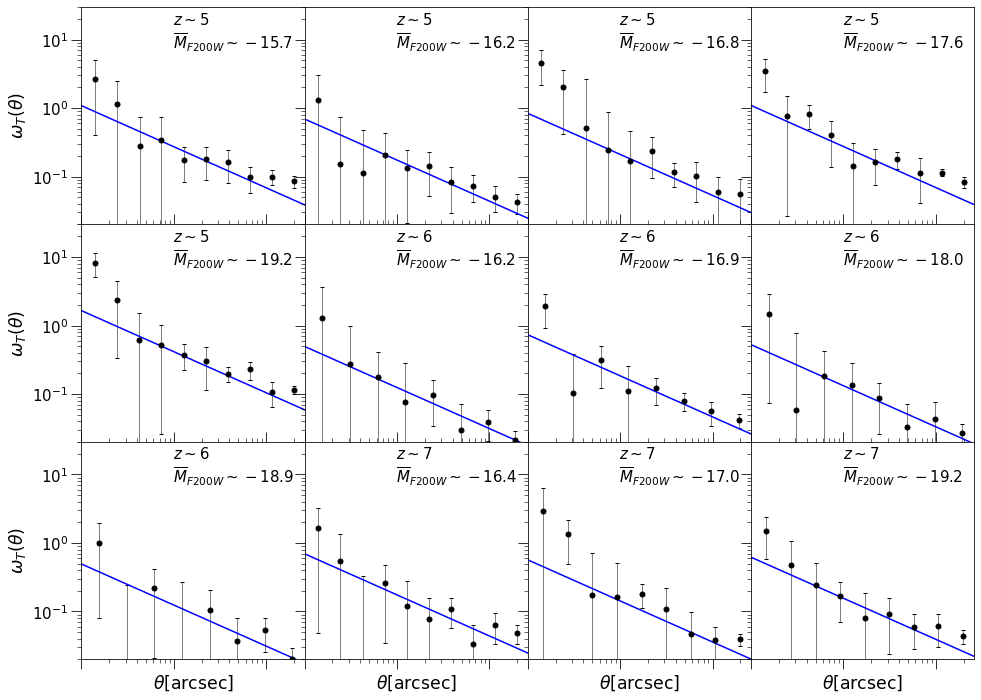}
\caption{ACFs measured in the GOODS-S field for each subsample are depicted in Fig.\ref{fig: bias vs M}. Solid lines represent the best-fit power-law function (Eq.\ref{eq: estimator model}) with $\beta=0.6$ held constant as in the main analysis. The mean redshift and absolute magnitude are indicated in the upper right corner of each panel.}
\label{fig: ACF multimag}
\end{figure*}

In this Appendix, we present as Fig.\ref{fig: ACF multimag} the detailed set of plots used in our analysis of the ACF measurements in order to investigate potential dependencies in galaxy bias evolution relative to the magnitude of the sample analyzed. The corresponding discussion and interpretation forms part of~Sec.~\ref{subsec: magnitude bias dependence}.

\bsp	
\label{lastpage}
\end{document}